# Oxide layer formation prevents deteriorating ion migration in thermoelectric $Cu_2Se$ during operation in air

Rasmus S. Christensen,[a] Peter S. Thorup,[a] Lasse R. Jørgensen,[a] Martin Roelsgaard,[a] Karl F. F. Fischer,[a] Ann-Christin Dippel,[b] Bo Brummerstedt Iversen*[a]

[a]Center for Integrated Materials Research, Department of Chemistry and iNANO, Aarhus University, DK-8000 Aarhus C, Denmark

[b]Deutsches Elektronen-Synchrotron DESY, Notkestraße 85, 22607 Hamburg, Germany

*Corresponding author: bo@chem.au.dk

**Abstract**

$Cu_2Se$ is a mixed ionic-electronic conductor with outstanding thermoelectric performance originally envisioned for space missions. Applications were discontinued due to material instability, where elemental Cu grows at the electrode interfaces during operation in vacuum. Here, we show that when $Cu_2Se$ is operating in air, formation of an oxide surface layer suppresses $Cu^+$ migration along the current direction. *In operando* X-ray scattering and electrical resistivity measurements quantify $Cu^+$ migration through refinement of atomic occupancies and phase composition analysis. Cu deposition can be prevented during operation in air, irrespective of a critical voltage, if the thermal gradient is applied along the current direction. Maximum entropy electron density analysis provides experimental evidence that $Cu^+$ migration pathways under thermal and electrical gradients differ substantially from equilibrium diffusion. The study establishes new promise for inexpensive sustainable $Cu_2Se$ in thermoelectric applications, and it underscores the importance of atomistic insight into materials during thermoelectric operating conditions.

**Main**

Over the past century, $Cu_2Se$ has been known as a promising thermoelectric (TE) material targeted *e.g.* for radioisotope thermal generators (RTGs) to supply energy in space.[1] However, the material remains unavailable commercially due to stability issues under operating conditions. The majority of studies characterizing the stability of $Cu_2Se$ have been performed *in vacuo*, however, for everyday applications characterisation at working conditions in atmospheric air are needed.[2]

The decomposition of $Cu_2Se$ in vacuum was described by Qiu *et al.* who found that material degradation depends on a critical electrical potential difference across the material pellet.[3] The study showed that under a thermal gradient Cu will migrate towards the cold junction and applying an external potential would increase $Cu^+$-migration and lead to decomposition.[3] Brown *et al.* observed decomposition at the hot junction when applying an opposing voltage, such that Cu is transported against the thermal diffusion.[1]

Common for these previous stability studies is that they are based on changes in macroscopic physical properties or *ex situ* characterisation methods. To fully understand the decomposition, it is preferable to characterise the structure and physical properties under working conditions. The Aarhus Thermoelectric Operando Setup (ATOS) allows for exactly this kind of characterisation by combining realistic device conditions with simultaneous collection of X-ray scattering and electrical transport data.[4, 5] ATOS has been used in studies of decomposition mechanisms in $Zn_4Sb_3$ and PbTe. [6, 7, 8, 9]

$Cu_2Se$ has been repeatedly rediscovered as a high performing TE material.[10] Recently, the interest in $Cu_2Se$ reignited with the development of the Phonon Liquid Electron Crystal (PLEC) concept.[11] PLEC materials exhibit a low lattice thermal conductivity resembling that of amorphous liquids together with crystal-like electronic conductivities. Liquid-like lattice thermal conductivity has been suggested especially in mixed ionic-electronic conductors such as $Cu_2Se$,[11] $Zn_4Sb_3$,[12] and $AgCrSe_2$,[13] although the exact origin of the low lattice thermal conductivity in these systems is still debated.

Liu *et al.* proposed that the liquid-like sublattice observed in $Cu_2Se$ inhibits the propagation of transverse phonons resulting in a reduced phonon mean free path.[11] However, Voneshen *et al.* showed, based on incoherent quasielastic neutron scattering (QENS), that the hopping timescales of Cu-ions are too slow to significantly affect lattice vibrations.[14] Instead, the ultralow thermal conductivity was attributed to strong anharmonicity. This was recently corroborated by a very high resolution single crystal X-ray diffraction study.[15]

$Cu_2Se$ is an intrinsic p-type semiconductor ($E_g$ = 1.23 eV).[16] It can be synthesised by melting,[17, 18, 19] ball milling,[20, 21] arc-melting,[22] or hydrothermal synthesis,[23, 24] and pellets can be produced with different compaction techniques. The disordered crystal structure of $Cu_2Se$ at room temperature (β-$Cu_2Se$) has been investigated since 1936,[25, 26] but was eventually solved in 2019 by Roth *et al.* using the 3D ΔPDF technique.[27] β-$Cu_2Se$ consists of two-dimensional ordered layers, which are stacked in a highly disordered fashion. A phase transition from the ion-insulating β-$Cu_2Se$ phase to the ion conducting α-$Cu_2Se$ is observed between 350 and 410 K depending on the degree of Cu-deficiency.[26, 28] The overall structure of α-$Cu_2Se$ can be described as a variation of the anti-fluorite structure with an ordered *fcc* arrangement of Se atoms and Cu disordered at positions at and around the tetrahedral holes. Often a simplified (and inadequate) split site model is used to describe the continuous Cu average electron density in analysis of X-ray diffraction data, where Cu is distributed on the tetrahedral 8c sites and on the 32f sites specifying the corners of the tetrahedra.[15] The position of the 32f site defines how far the corners of the tetrahedra extends around the 8c site.[26] Based on QENS it was proposed by Danilkin *et al.* that this displacement of Cu is directly linked to the ion conducting properties.[29] X-ray and neutron diffraction has indicated that the diffusion of Cu-ions under equilibrium conditions occurs between the tetrahedral sites along the <111> direction, while avoiding the octahedral site.[15, 30, 31] These previous studies of the Cu substructure were all performed at a constant uniform temperature, hence ignoring the fact that a thermal gradient or external applied current may impact the Cu migration pathways in the material.

Here we show that in air, the formation of a $Cu_2O$ surface layer significantly affects the decomposition of $Cu_2Se$. The formation of $Cu_2O$ effectively alters the Cu-migration mechanism proposed by Qiu *et al.*,[3] while maintaining the high performing thermoelectric properties. We also visualise the three-dimensional Cu substructure and ion migration pathways while applying a thermal gradient and current.

**Results**

Three *in operando* experiments with the ATOS were performed on $Cu_2Se$ pellets using a thermal gradient of 750 K to 300 K and applying electrical current densities of 0, 0.5 and 1.0 A/mm$^2$; hereafter referred to as CS00, CS05 and CS10, respectively. The CS10 experiment shows the β- to α-$Cu_2Se$ phase transition after ~10 min. seen in Fig. 1a. After the phase transition no large structural changes are observed in the *in operando* diffractograms, *i.e.* no decomposition of $Cu_2Se$ or no new phases are appearing with time. The only impurity observed alongside the α-$Cu_2Se$ phase is $Cu_2O$ as shown in Fig. S1. This impurity is observed in all three experiments.

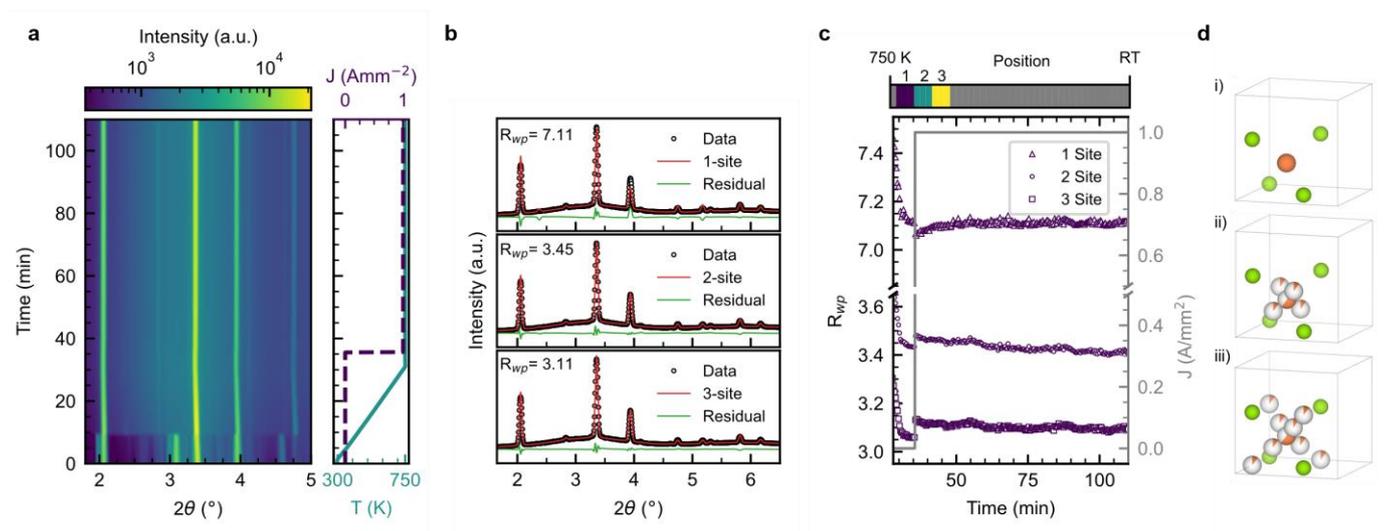

**Figure 1** a) 2D contour plot of CS10 position 1 together with an overview of the experimental conditions. b) Rietveld refinement of the frame at position 1 after 50 minutes for 1-site, 2-site, and 3-site models, respectively. c) $R_{wp}$ plotted as a function of time for the 1-site, 2-site, and 3-site models, respectively. d) Illustration of the structure used in the i) 1-site, ii) 2-site, and iii) 3-site model.

Rietveld refinements were performed using three different two-phase models ($Cu_2Se$ and $Cu_2O$), where the Cu positions in $Cu_2Se$ were distributed on 1-site (8c), 2-sites (8c and 32f) or 3-sites (8c, 32f and 4b) illustrated in Fig. 1d. The models are compared in Fig. 1b-c, and the agreement factor $R_{wp}$ decreases from 7.11% for the model with one Cu site to 3.11% for the 3-site model. The absence of the (200) reflection at 2.3° is best described by the 3-site model indicating that Cu occupies sites very close to or in the octahedral site. The 3-site model was used for all the following sequential refinements.

For CS10 thermal equilibrium is assumed after approximately 50 min based on the obtained unit cell and measured electrical resistance (Fig. S2). The Cu substructure can be visualised by calculating the Maximum Entropy Method (MEM) density at position 1 after 50 min. (Fig. 2a). The MEM density shows the tetrahedrally shaped Cu structure similar to the one found by Roth *et al.*[15] In addition, the MEM density shows significant density (~ 3 electrons/Å$^3$) in the octahedral site at (0.5, 0.5, 0.5), which further justifies the choice of model for the sequential refinements. The presence of electron density in the octahedral site is counter to previous studies, and it can be attributed to the electrical current and thermal gradient resulting in different Cu-migration pathways compared with isothermal characterisation. The fact that migration is possible through the octahedral site indicates that ion-migration occurs in a 3D network under thermoelectric operating conditions.

The refined total Cu-occupancy, Fig. 2b, shows a significant decrease at position 1 (~1%) and little to no decrease at position 2 and 3. The decrease in Cu-occupancy of the $Cu_2Se$ phase can be ascribed to two contributions, namely $Cu^+$-migration and $Cu_2O$ formation at the surface. In the event of $Cu^+$-migration the refined change in occupancy describes the net $Cu^+$-ion flux within the volume defined by the X-ray beam. The observed flux is different at each position which signifies a large temperature dependence of the ion migration *i.e.* the Cu-ions are more mobile near the hot junction as expected.[8] The formation of $Cu_2O$ at the surface is observed as a secondary crystalline phase and the refined weight percentage is shown in Fig. 2c for CS10 and Fig. S3 for all three experiments. The amount of Cu$_2$O observed can be directly related to a change in occupancy by assuming the following reaction.

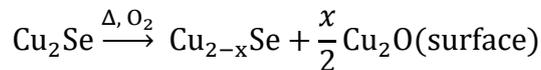

$$Cu_2Se \xrightarrow{\Delta,\ O_2} Cu_{2-x}Se + \frac{x}{2} Cu_2O(\text{surface})$$

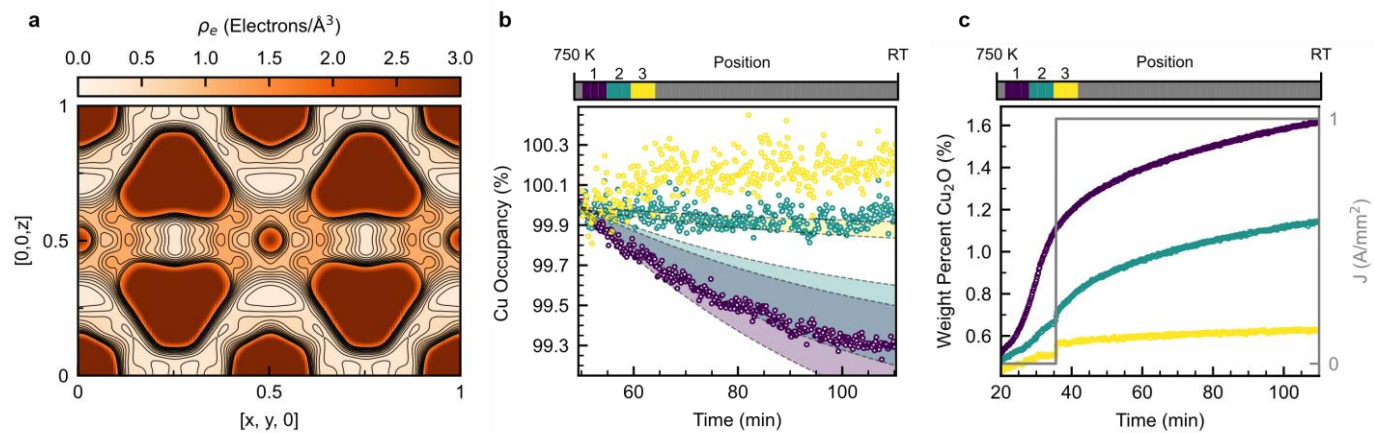

**Figure 2. a**) MEM density of Cu$_2$Se after approx. 50 min. Black contour lines are plotted in steps of 0.1 from 0.1 to 1.5 electrons/Å$^3$. **b**) Change in total Cu occupancy after assumed thermal equilibrium for CS10. The colour code designates measurement positions on the pellet. The dashed lines and shaded area correspond to the expected change in total Cu occupancy as estimated from the Cu$_2$O formation. Top and bottom of the span corresponds to assuming all or half of the formed Cu$_2$O is probed by the X-ray beam, respectively. **c**) Observed change in weight percent of Cu$_2$O with time for CS10.

As the dimensions of the bar are larger than the height of the incident X-ray beam the obtained Cu$_2$O weight percent is underestimated (as illustrated in Fig. S4). In Fig. 2b the estimated occupancy change based on the refined Cu$_2$O weight percent is shown as a coloured region. The upper limit of the region corresponds to assuming that the probed volume contains all the formed Cu$_2$O, while the lower limit corresponds to assuming that the probed volume only contains half of the formed Cu$_2$O. The observed Cu$_2$O weight fraction describes the refined occupancy well for position 1 whereas it is overestimated for position 2 and 3. The fact that the observed occupancy change in position 1 is described well by the amount of Cu$_2$O observed indicates that the net Cu$^+$-ion flux in and out of the probed volume is near zero. The overestimation observed in position 2 and 3 indicates that the net flux of Cu$^+$-ions into the probe volume is larger than the flux out of the volume. The different changes in occupancies at different positions indicate that under working conditions the formation of $Cu_2O$ limits the extent of $Cu^+$-ion migration. A similar analysis was performed for CS00 and CS05 as shown in Fig. S5, but here the reduced data quality makes the trends less clear.

In the CS10 experiment the applied current density was five times greater than the critical current density proposed by Qiu *et al.*,[3] but no change in the crystal structure is observed at the cold junction. This shows that the additional force directed towards the surface caused by oxide formation is competing with the directional migration force of the applied electric field. The oxide formation dominates, and it inhibits a build-up of $Cu^+$-ions at the cathode. The assumptions of critical voltage therefore only refer to specific measurements in vacuum. The study by Qiu *et al.* found that the critical chemical potential difference of Cu defines the point of decomposition at the cathode (corresponding to J = 18-21 $Acm^{-2}$).[3] However, as shown in Fig. S1A and Fig. S3A, $Cu_2O$ forms even without an applied current (the CS00 experiment). The presence of $O_2$ instead of vacuum alters the Cu-migration mechanism, and the formation of $Cu_2O$ is favoured relative to migration. The change in occupancy and the refined $Cu_2O$ weight percent (see Fig. 2) indicates that the system approaches a steady state where no more Cu is moving with the current *i.e.* the system is approaching a lower limit of the Cu occupancy. This limit could be a result of negligible Se evaporation expected considering the Se vapour pressure of 0.03 bar at 750 K.[32] For operation in vacuum however, Se evaporation has been identified as a cause for severe stability issues.[1]

Fig. 3a shows the weight percent of the high temperature α-$Cu_2Se$ phase upon cooling for CS10. Comparing the different positions, a change in composition is observed at the positions closest to the actively heated end. The weight percent of α-$Cu_2Se$ at 350 K, *i.e.* below the phase transition temperature, remains as high as ~43 % in the actively heated end. The stabilisation of the α-$Cu_2Se$ phase at low temperature is caused by a Cu deficient lattice.[28] The Cu deficiency in the lattice can be directly related to the formation of $Cu_2O$ by comparing the weight percent of $Cu_2O$ before cooling and the final weight percent of α-Cu$_2$Se. A linear relation between Cu$_2$O and α-Cu$_2$Se is observed (Fig. 3b), which shows that the formation of Cu$_2$O is directly coupled to the stability under operating conditions. X-ray diffraction data along the bars of the CS10 and CS05 samples after the *in operando* measurements show significant stabilisation of α-Cu$_2$Se at room temperature (Fig. S6), which according to the phase diagram proposed by Ogorelec *et al.* corresponds to a composition of approx. Cu$_{1.80}$Se.[33]

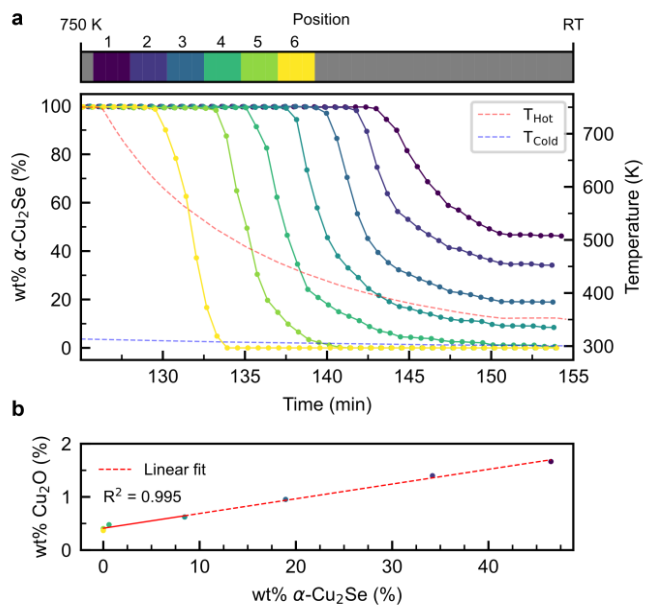

**Figure 3.** a) Refined weight percent of α-Cu$_2$Se during cooling for CS10. The red and blue dashed lines mark the measured temperatures of the hot and cold junction, respectively. b) Weight percent of Cu$_2$O immediately before cooling versus the final weight percent of α-Cu$_2$Se for different positions in the sample. The red dashed line is a linear fit.

Measurements of the temperature dependent Seebeck coefficient and electrical resistivity of the as synthesised sample reveal comparable properties to those reported by Liu *et al.*, and this shows that indeed high performance Cu$_2$Se was synthesised by the present SPS method (Fig. S7).[11] Thermoelectric properties after the *in operando* experiment were measured on CS05, shown in Fig. 4. The electrical resistivity and Seebeck coefficient are lower at all temperatures resulting in a lower power factor at room temperature. However, above the β → α-Cu$_2$Se phase transition, the power factor is virtually unchanged from that of the as-synthesised sample. The fact that the power factor is nearly unaffected by the *in* *operando* measurements and the formation of $Cu_2O$ means that the *zT*-value should remain high as well. The exact value will depend on the thermal conductivity, which we unfortunately cannot measure on these samples. The electronic contribution to the thermal conductivity is expected to increase due to the nearly halved electrical resistivity and higher Lorenz number indicated by the reduced Seebeck coefficient.[34] Changes in the lattice contribution are harder to predict, but it is unlikely to decrease by much, since the thermal conductivity of $Cu_2Se$ is already very low.

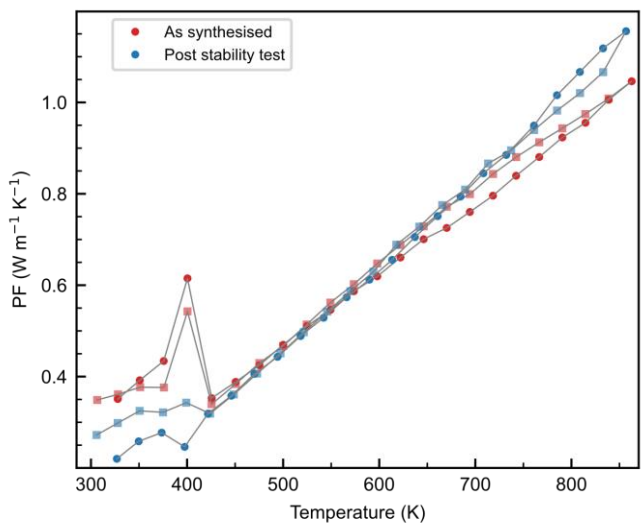

**Figure 4.** Power factor, $S^2\sigma$, for the as-synthesised $Cu_2Se$ sample(red) and the CS05 sample after the *in operando* measurements (blue). Data points collected during heating and subsequent cooling are plotted as circles and squares, respectively.

The CS05 sample was examined using SEM-EDX after the *in operando* stability test. The striations observed in the SEM images is caused by polishing. A region close to the actively heated end is shown in Fig. 5. A layer of ~5 μm thickness is found on the surface of the bar. The layer appears bright in the SEM image indicating a compositional difference from the darker bulk $Cu_2Se$. The corresponding EDX analysis shows that the layer primarily consists of Cu and O. The Cu and Se in the bulk $Cu_2Se$ appears to be homogeneously distributed. The oxide layer observed by SEM-EDX is exclusively found on the surface, and only in regions close to the hot end heated to 750 K (see also Fig. S8 and S9). This corresponds well with the results from the *in operando* X-ray scattering experiment where the refined Cu$_2$O weight percent is higher closer to the hot end.

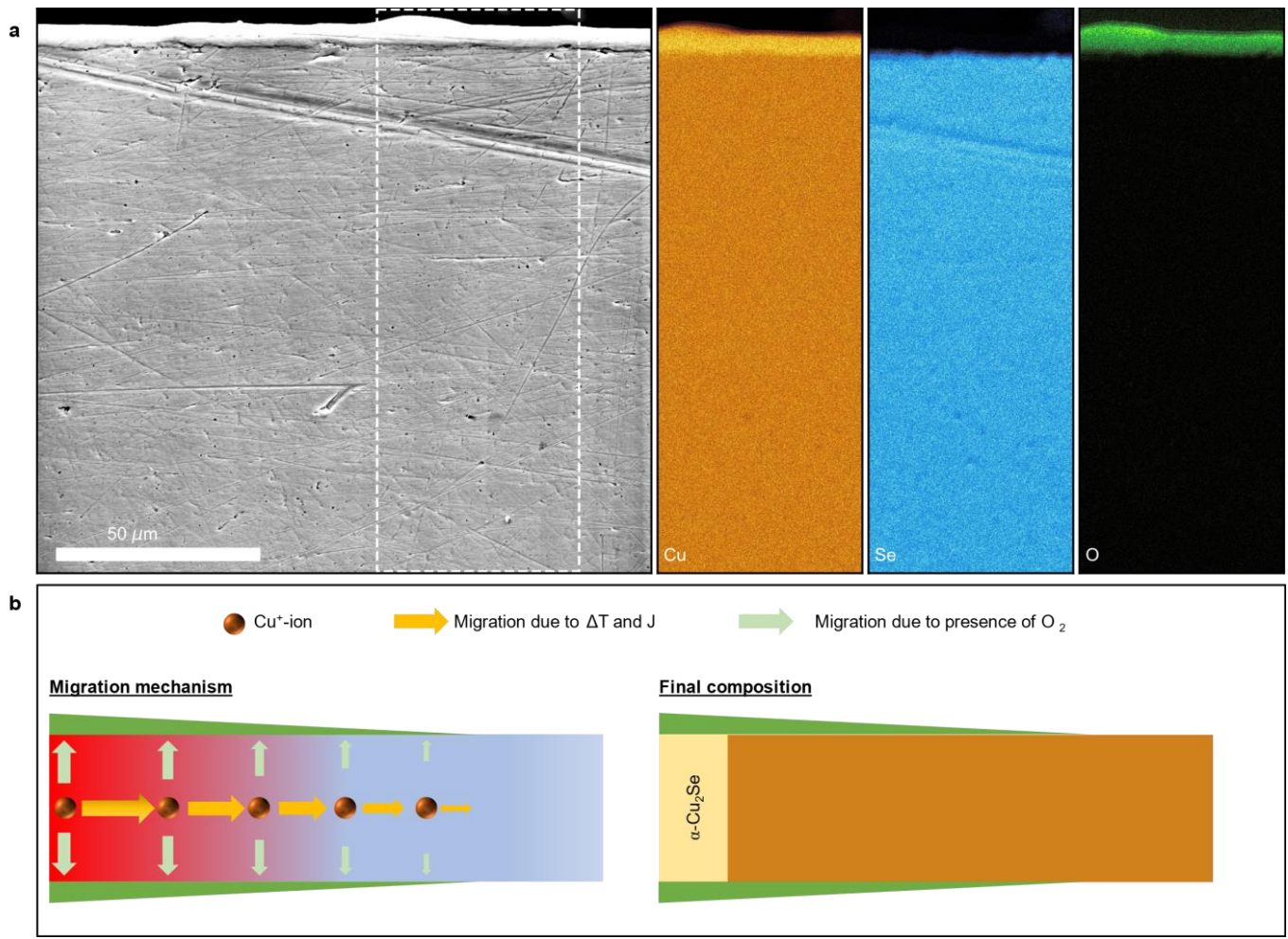

**Figure 5. a**) SEM image of CS05 after the *in operando* stability test near the hot end and EDX-maps of Cu, Se and O in the rectangle marked by the white dashed lines. **b**) Illustration of the proposed Cu migration mechanism with the green areas representing Cu$_2$O formation.

Based on the *in operando* X-ray diffraction data and SEM analysis an illustration of the Cu migration mechanism in the presence of O$_2$ is proposed in Fig. 5b. The sketch shows the Cu$^+$-ion migration along the thermal gradient and the applied current (yellow arrows), and also towards the surface (green arrows) due to the chemical drive to form Cu$_2$O. The oxidation effectively limits the migration along the thermal gradient and current, and consequently prevents the degradation at the cathode. However, as Cu leaves the Cu$_2$Se lattice to form Cu$_2$O, a gradual change in stoichiometry is found upon cooling to room temperature illustrated as a region of remnant α-$Cu_2Se$. This is an important result to consider when attempting to further stabilise $Cu_2Se$ using segmentation, surface coating, or similar methods as the formation of $Cu_2O$ is highly temperature dependent and occurs even when no current is flowing. The formation of $Cu_2O$ is expected to be independent of segmentation and further stabilisation using segmentation is not expected. However, the *in operando* stability tests in air show that the formation of $Cu_2O$ limits ion migration and thus has a stabilising effect similar to segmentation. Attempting to stabilise the material using surface coating would affect the amount of $O_2$ available hence resembling vacuum conditions and thus influence the migration of Cu.

An *ex situ* (*i.e.* without X-ray beam) stability test was conducted in the ATOS with the current applied opposite to that of the thermal gradient. The thermal gradient was the same as for the *in operando* experiments and the applied current density was 0.5 A/mm$^2$ (Fig. S10). A SEM-EDX image of the actively heated end is shown in Fig. 6. The SEM image shows that a ~80 μm large protrusion has formed, which is similar to the Cu "nanowire bundles" observed in the study by Brown *et al.*[1] In their study, Cu deposition was also only observed at the actively heated end, when the current was applied in the opposite direction of the thermal gradient. The protrusion observed in the present *ex situ* experiment consists primarily of Cu as can be seen from the EDX map in Fig. 6. There is a uniform oxygen content since this bar was not polished after the stability test. The SEM-EDX maps also show that the Cu and Se content is no longer uniformly distributed, which indicates that there are regions where the $Cu_2Se$ lattice has decomposed. The fact that Cu deposition is observed with a reverse current agrees well with the proposed migration mechanism in Fig. 5b. The total $Cu^+$-ion flux is now directed towards the hot end of the bar, as the flux from the electrical current overpowers the flux from the thermal gradient. This can be explained by the more mobile $Cu^+$-ions at elevated temperatures, which was also observed in the *in operando* scattering experiments. Therefore, a Cu-rich area is formed at the junction between the heat source and the sample. In this junction there is little $O_2$ available, which then promotes deposition of metallic Cu rather than oxidation.

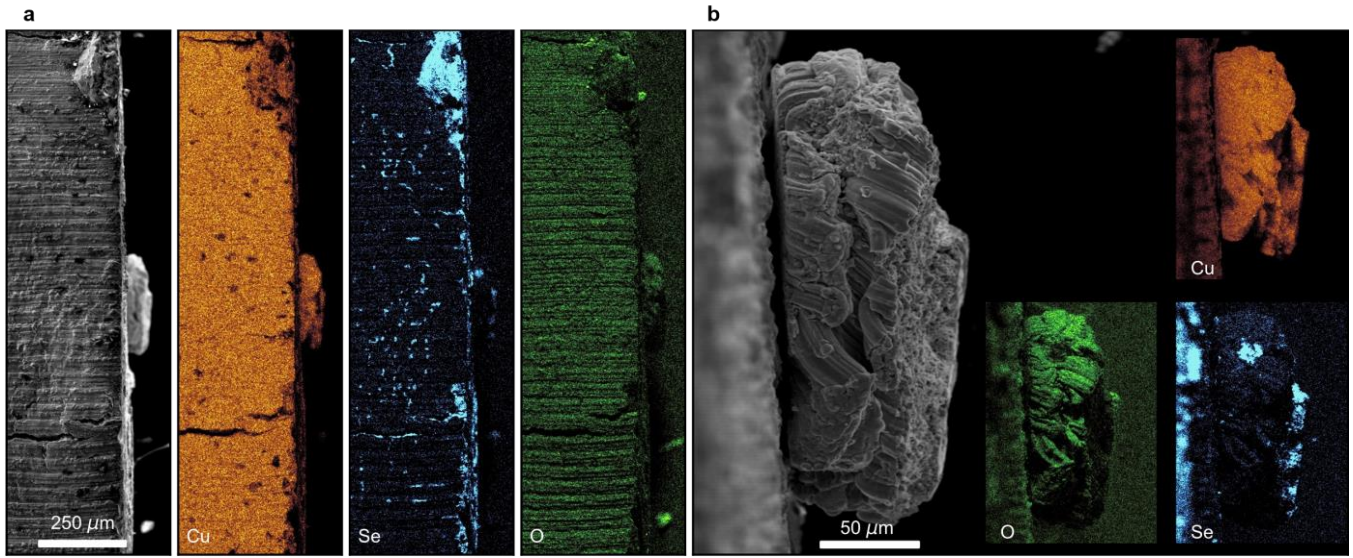

**Figure 6.** SEM image collected with secondary electrons of the *ex situ* sample after stability test. **a**) Wide focus of the actively heated end. **b**) Focus on the observed protrusion in the actively heated end. EDX maps for Cu, Se, and O is shown for each SEM image.

In summary, the stability of $Cu_2Se$ in air was studied using *in operando* X-ray scattering in combination with simultaneous measurements of electrical resistivity. No structural decomposition was observed despite applying a thermal gradient of 750 K to 300 K and a current density of up to 1 A/mm$^2$. The Cu-migration pathway was visualised using a MEM density map obtained while applying a temperature gradient and external current. Significant density was observed in the octahedral hole showing a 3D migration different from the pathway observed at equilibrium conditions. A decrease in Cu occupancy was observed directly related to the amount of $Cu_2O$ being formed at the surface. The presence of $O_2$ affects the ion migration, as the chemical drive for formation of $Cu_2O$ is more energetically favourable than expulsion of elemental Cu. This in turn inhibits the pathways that in vacuum lead to structural decomposition. Upon cooling, a Cu-deficient lattice was observed and the α-$Cu_2Se$ phase was stabilised at room temperature. The Cu-deficiency was found to be linearly related to the observed $Cu_2O$ weight-percent.

Transport property measurements showed that the thermoelectric power factor above the α→β phase transition was virtually unchanged during the *in operando* stability test, despite the formation of $Cu_2O$. SEM-EDX analysis showed the hot end to have a homogeneous distribution of Cu and Se in the bulk $Cu_2Se$ and a ~5 μm thick $Cu_2O$ layer. Combining the *in operando* X-ray scattering and SEM-EDX led to the simple description of the Cu-migration mechanism as shown in Fig. 5b.

This study shows that careful consideration is necessary when preparing a thermoelectric module for application in air using $Cu_2Se$. Especially the directions of the applied thermal gradient and electric current are important to keep in mind as they can affect the stability significantly. Previous conclusions based on applications in vacuum are shown to be invalid for samples exposed to air. In particular, the critical voltage defined for formation of elemental Cu at the interfaces is not valid since this effect is overwhelmed by the chemical drive to form $Cu_2O$. The formation of $Cu_2O$ reaches a steady state, where the thermoelectric properties remain very high, and thus the formation of the $Cu_2O$ surface layer turns out to be advantageous for application of $Cu_2Se$ in thermoelectric modules. Finally, the formation of $Cu_2O$ does not affect the electrical contacts since there is no oxygen present at those interfaces.

**Acknowledgements**

This work was supported by the Villum Foundation. We gratefully acknowledge PETRA III, Deutsches Elektronen-Synchrotron, DESY, a member of the Helmholtz Association, for the provision of experimental facilities and we thank the beamline staff for support. Parts of this research were carried out at beamline P21.1 with beamtime allocated for proposal I-20191208, I-20200441 and I-20210475. Andreas Dueholm Bertelsen and Kasper Andersen Borup are thanked for their support during the beamtimes. Espen Drath Bøjesen is thanked for help with the SEM-EDX measurements and interpretation. We thank the Carlsberg foundation for funding the Clara SEM.

**Method**

**Synthesis**

Stoichiometric amounts of $Cu_{2.05}Se$ were ground in an agate mortar and cold pressed into a ¼ inch pellet. The cold pressed pellet was sealed in a quartz ampoule with a vacuum and pre-synthesised at 773 K for 24 hours after which the temperature was raised to 1423 K for 24 hours. Upon cooling the sample was further annealed at 723 K for 48 hours. Unreacted Cu was removed before grinding in an agate mortar and further sieving to a particle size below 63 μm. The powder was then compacted using spark plasma sintering (SPS-515 Syntex Inc., Japan) at 873 K using 60 MPa. Potential-Seebeck-Microprobe analysis show that the synthesised pellet is homogeneous (Fig. S11).[35] Powder X-ray diffraction (PXRD) data was obtained prior to SPS pressing using an in-house Rigaku Smartlab diffractometer equipped with a Cu source (See Fig. S12).

*In operando* **X-ray experiment**

High energy X-ray diffraction experiments were performed at beamline P21.1 at PETRA III, DESY, Germany. The diffraction data were collected in transmission geometry, which is possible due to the high X-ray energy of 101.7 keV. Calibration of the sample-to-detector distance as well as the instrumental resolution was performed using a NIST 660b $LaB_6$ line broadening standard and PyFAI-Calib2.[36] The scattering data was collected using a 0.8 x 0.8 $mm^2$ beam profile with a PILATUS3 X CdTe 2M detector centred approximately 750 mm from the sample yielding a maximal scattering vector in Q of ~10 $Å^{-1}$. The data were collected with an acquisition time of 3 s. The sample was moved continuously between different sample positions using the heavy load diffractometer. The resulting diffraction patterns were azimuthally integrated using PyFAI and normalised with respect to an intensity monitor.[36]

Three *in operando* experiments were performed using ATOS, which has been described in detail by Jørgensen *et al*.[5] All experiments were performed using a thermal gradient of 750 K to 300 K and applying electrical current densities of 0, 0.5 and 1.0 $A/mm^2$; referred to as CS00, CS05 and CS10, respectively. A schematic overview of each experiment is shown in Fig. S13.

The diffraction patterns were refined sequentially using Topas-Academic V6.[37] The α- and β-$Cu_2Se$ phases were modelled using the average crystal structures proposed by Eikeland *et al.* (ICSD 243957, 243953)[26] and the $Cu_2O$ phase was modelled using the crystal structure proposed by Restori *et al.* (ICSD 63281).[38]

The *in operando* data was modelled using different two-phase models (α-$Cu_2Se$ and $Cu_2O$) for all datasets at temperatures above the phase transition temperature. Three different models were tested for the α-$Cu_2Se$ structure with one, two or three Cu sites. The composition was assumed to be $Cu_2Se$ in all three models. For the 1-site model 100 % Cu were placed in the tetrahedral 8c site. For the 2-site model 60 % Cu were placed in the tetrahedral 8c site and the remaining 40 % were placed in the 32f site. In the 3-site model 1.5 % Cu were placed in the octahedral site, 59.5 % Cu were placed in the tetrahedral 8c site and the remaining 39% were placed in the 32f site. Different 3-site models were tested, before settling on this particular model. The refined parameters were scale, unit cell parameters, and a background. The background consisted of 3 broad Gaussian peaks together with a 5$^{th}$ degree Chebyshev polynomial. For the $Cu_2O$ phase the atomic displacement parameters (ADPs) were kept fixed due to a lack of peaks at large scattering angle. For the α-$Cu_2Se$ phase two separate refinements were done. Initially, the individual ADPs were refined freely for the entire dataset with the occupancies fixed. Subsequently, the ADPs were fixed to the values obtained after 50 minutes and then the occupancy of Cu was refined freely.

Refinements of cooling experiments and room temperature data were performed using a two-phases model with the α- and β-$Cu_2Se$ phase together. Again, the scale factor, unit cell and a background were refined as well as the ADPs for Cu and Se.

The peak profile was modelled using the instrumental resolution function as well as an additional Lorentzian contribution (the X parameter). As the sample is larger than the 0.8 x 0.8 mm$^2$ beam profile the Thompson-Cox-Hastings pseudo-Voight peak profile was convoluted with a top hat function with a fixed length found based on the LaB$_6$ NIST line broadening standard equal to the *Source_Width* macro in Topas V6.[37]

The MEM calculation was done using the Sakato-Sato algorithm in the BayMEM program.[39, 40] A grid of 140x140x140 voxels was used while assuming flat prior density. The convergence criteria were obtained at $\chi^2 = 1$.

**Thermoelectric properties**

The electrical resistivity and Seebeck coefficient were measured before and after the *in operando* stability tests using an Ulvac ZEM-3. Measurements were performed between 298 K and 873 K in steps of 25 K and heater temperature differences of 5, 10, 15, 20, 25, and 30 K were used for the dV/dT data to extract the Seebeck coefficient using the slope method.

**Scanning electron microscopy**

Scanning electron microscopy (SEM) images were collected using a TESCAN CLARA SEM. The secondary electron images were collected using an E-T detector whereas a 4-quadron detector was used for images recorded using backscattered electrons. An electron beam energy of 20 keV and current of 3 nA was used for all images. The CS05 sample was finely polished using 1/4 μm grain diamond paste. No polishing was done for the *ex situ* sample to avoid breaking the sample.

**Data availability**

Data sets generated during the current study are available from the corresponding author on reasonable request.